\documentclass[12pt]{article}
\usepackage{xcolor}
\usepackage{graphicx}
\begin{document}

\vspace*{1cm}
\thispagestyle{empty}
\centerline{\large\bf Hydrogen Atom: Its Spectrum and Degeneracy}
\vspace*{.5cm}
\centerline{\bf Importance of the Laplace-Runge-Lenz Vector} 
\bigskip
\begin{center}
Akshay Pal \footnote{Email:ug18ap@iacs.res.in,akshayphysics1804@gmail.com}\\
{\em  Department of Theoretical Physics, IACS, Kolkata,India}\\
Siddhartha Sen\footnote{Email: sen1941@gmail.com}\\
{\em Centre for Research in Adaptive Nanostructures and Nanodevices},\\
{\em School of Physics, Trinity College Dublin, Ireland}\\

\end{center}
\vskip.5cm
\subsubsection*{Abstract}
Consider the problem: why does the bound state spectrum $E(n)<0$, of hydrogen atom Hamiltonian $H$ have more degenerate eigenstates than those required by rotational symmetry? The answer is well known and was demonstrated by Pauli. It
is due to an additional conserved vector, $\vec{A}$, of $H$, called the Laplace-Runge-Lenz vector, that was  first discovered for planetary orbits.  However, surprisingly, a direct link between degenerate eigenstates of $H$  and the physical labels that describe them is missing. To  provide such a link requires, as we show, solving a subtle problem of self adjoint operators. In our discussions we address a number of conceptual historical aspects regarding hydrogen atom that also include a careful discussion of both the classical as well as the quantum vector $\vec{A}$.
\newpage

\subsubsection*{Introduction}
Our aim is to understand the quantum degeneracy of eigenstates of  the hydrogen atom Hamiltonian $H$ in terms of its two conserved vectors, the conserved Laplace-Runge-Lenz vector $\vec{A}$ and the angular momentum vector $\vec{L}$.  On the way we discuss a number of historical conceptual problem.

It is well known that the bound state spectrum of atomic  hydrogen $E(n)$, labeled by the principal quantum number $n$ is independent,of both the angular momentum $l$ related to the eigenvalue of the operator $L^2=\vec{L}.\vec{L}$ and any one component of $\vec{L}$ chosen to be $L_z$ with eigenvalue $l_z$,  that forms part of the complete set of commuting operator, $(H,L^2,L_z)$ whose eigenvalues label an eigenstate of $H$. For the principal l quantum number $n$, the labels $(l,l_z)$ can
range over the values  $(0\leq l\leq (n-1))$ and $(-l\leq l_z \leq +l)$ respectively. 
The $l_z$ degeneracy comes from the rotational invariance of $H$ while the $l$ degeneracy is thought to be due to the presence of the conserved vector $\vec{A}$.

These features were  established by Pauli\cite{pauli} almost immediately  after Heisenberg formulated quantum mechanics as an operator theory, where the ideas of eigenvalues and eigenstates first appeared.  Pauli\cite{pauli} used the operator  approach of  Heisenberg  to show how both the energy eigenvalues and the number of degenerate eigenstates that they have could be determined by an algebraic method simply from the presence of the two conserved vector operators $(\vec{L},\vec{A})$ of
the Hamiltonian operator $H$ for hydrogen atom. Thus a direct link between the symmetries of $H$ and degeneracy of states was established.
  
 We will outline Pauli's calculation and show that although his approach links degeneracy of states to symmetries of $H$ it  cannot be used to find the eigenvectors of $H$ neither can it properly identify the physical nature of the degeneracy found by calculation.  Thus although the vectors $(\vec{L},\vec{A})$ are clearly related to the degeneracy problem of hydrogen a direct physically transparent link is missing. 
    
After outlining Pauli's method and pointing out its shortcoming, we proceed to resolve 
the issues raised. The first step in this direction we take is to solve  Schroedinger's equation, using an algebraic factorization method, that allows us to find the eigenvalues and eigenvectors. In this approach explicitly establishes and uses $l$ degeneracy of the energy spectrum and it also give an intuitive physical picture of the principal quantum number.  But in this approach the vector $\vec{A}$ plays no role. To remedy this gap we proceed to express  $\vec{A}$, in terms of the hydrogen atom eigenstate quantum numbers. Our aim is to investigate if such a representation allows us to understand the degeneracy of eigenstates revealed by calculations. To our surprise we  do this, we run into a problem. We find that the eigenvalues of $A^2=\vec{A}.\vec{A}$ can be negative: the operator $\vec{A}$ is thus not self adjoint in the space of the eigenfunctions of hydrogen. However, in our discussion of Pauli's approach, we show $A^2$ is always positive and self adjoint in the space of eigenfunctions considered by Pauli. There is thus a subtle problem. We show how  the problem can be resolved by an extension of $\vec{A}$, which we call $\vec{B}$, that is also a conserved vector of $H$.  Then by taking $B^2=\vec{B}.\vec{B}$,  as an additional quantum label for the eigenstates of hydrogen, we are able to explain the degeneracy of eigenstates and directly link the degeneracy to the conserved vectors of $H$. 

Since the problem discussed is historical, for completeness, we sketch well known ideas such as the classical instability of  electron orbits, state Bohr's ad hoc resolution of the problem and its subsequent resolution by quantum mechanics. We also determine the the properties of $\vec{A}$ for the classical and quantum case that we need to sketch Pauli's calculation of the hydrogen bound state  spectrum followed by the equivalent Schroedinger equation method presented using the algebraic factorization method. We then establish our results. 
 
\subsubsection*{Problem of a classical orbit model and Bohr's Quantum Model}
We start at the beginning.  Balmer's formula for the spectral lines of hydrogen is best expressed using the inverse of wavelength of the lines as the difference between two terms of the form $\frac{R}{n^2}, n=1,2,...$ with $R=109737 cm^{-1}$. To understand this result a first step is to establish an model for hydrogen atom. This was done by Rutherford, who discovery from his scattering experiments, that hydrogen atom could be pictured as a negatively charged electron orbiting a localized positively charged proton, rather than a diffuse positive charge with a negatively charged electron embedded in it. The experimental results suggested a picture for hydrogen atom as an   orbiting electron round the proton. However it is well known that such a classical orbiting picture is unstable. The orbiting electron accelerates as it moves along a circular or elliptic orbit, consequently, from Maxwell's laws, it must radiate electromagnetic waves, loosing  energy to eventually collapse to the nucleus. 

Let us estimate the classical lifetime of this collapse, by considering an electron in a circular orbit of radius $a_0=\frac{\hbar^2}{me^2}$, the lowest Bohr radius, and determine the  time it takes for such an electron to collapses to the center, where $r=0$ following a series of circular orbits of decreasing radii. To do this we need to define the energy of the orbiting electron and the  rate at which it looses energy by radiating  electromagnetic waves.
 
 The energy $E$ of an electron of velocity $\vec{v}$, charge $e$  and mass $m$ located at a point $\vec{r}$ is, 
 \begin{displaymath}
 E=\frac{1}{2}mv^2-\frac{e^2}{r}=-\frac{e^2}{2r^2}~~\mbox{using Newton's law}
 \end{displaymath}
 where $r=+\sqrt{\vec{r}.\vec{r}}$. We have assumed that the electron is moving in a circular  orbit of radius $r$. The energy loss of such an electron per second can be written in two ways. First from the orbit energy it follows  that,
 \begin{displaymath}
 \frac{dE}{dt}=+\frac {e^2}{2r^2}\frac{dr}{dt}
 \end{displaymath}
 This expression relates energy change to orbit size change while the second
 expression relates energy change to due to electromagnetic  radiation, given by\cite{jackson},
 \begin{displaymath}
 \frac{dE}{dt}=-\frac{e^2a^2}{3c^3}=-\frac{2r_0}{3r^4}r_0c^3
 \end{displaymath}
 where $a^2$, the square of the orbiting electron acceleration and we have used Newton's law for a circular orbit to write  $a^2=\frac{v^4}{r^2}=\frac{e^4}{m^2r^4}$. 
 
From these two equations it follows that $r^2\frac{dr}{dt}=-\frac{4}{3}r_0^2c^3$ where  $r_0=\frac{e^2}{mc^2}$, and $c$ is the velocity of light. Integrating this equation we get
  \begin{displaymath}
 r^3=a_0^3-4r_0^2c~t 
 \end{displaymath} 
 where we have set $r=a_0$ at time $t=0$. The fall to center time is then $T=\frac{a^3_0}{4r^2_0c}\approx  10^{-9}$ seconds, taking $a_0\approx 10^{-8}$ cm, the Bohr radius and $r_0=\frac{e^2}{mc^2}\approx 10^{-13}$ cm. 
\subsubsection*{Bohr's resolution of the instability problem} 
Thus a classical orbiting charged particle is highly unstable and must collapse to the center. It cannot be a  classical model for hydrogen atom.. Bohr resolved the problem in a radical way, by introducing postulates that disregarded the laws of Maxwell. He extended the idea of quanta, introduced by Planck for energy, to angular momentum and suggested that radiation of electromagnetic waves only occurs when there are transitions between postulated orbits of different energies in which an electron violates Maxwell's laws and does not radiate.  These special orbits, stationary state energy levels, were postulated to have integral units of angular momentum and it was shown that electrons in these orbits had energies $E_n=-\frac{\alpha^2mc^2}{2n^2}$ , where $\alpha=\frac{e^2}{\hbar c}$,was the fine structure constant, $m$ the mass of the electron, $c$ the speed of light in vacuum. From this expression the value of $R$ in Balmer's formula could be determined, since $\frac{E_n}{\hbar c}=\frac{1}{\lambda_n}=\frac{R}{n^2}$. For the sake of completeness we sketch Bohr's calculation. 

The starting point was to model hydrogen atom as an electron of mass $m$ and charge $-e$ moving in a circular orbit of radius $r$ where a proton of charge $+e$ was at the center of the orbit. Neglecting the motion of the proton, the equation of the motion and energy $E$  of the electron are,
\begin{eqnarray}
m\frac{d^2 x}{dt^2}&=&\frac{e^2}{r^2}=\frac{mv^2}{r}\\
mv^2&=&-\frac{e^2}{r}\\
E&=&\frac{mv^2}{2}-\frac{e^2}{r}
\end{eqnarray}
Bohr's two postulates were
\begin{eqnarray}
mvr&=&n\hbar, n=1,2,.. ~~\mbox{stationary state condition}\\
\hbar\omega&=&E_n-E_m~~\mbox{photon emission condition}
\end{eqnarray}
The equation of motion shows that $\frac{mv^2}{2}=\frac{e^2}{r}$, so that   $E=-\frac{e^2}{2r}$, while using these results in the square of the stationary state condition fixes the value of $r$ for given value for $n$ value, giving the result, $E_n=-\frac{\alpha^2mc^2}{2n^2}$. 

A conceptual puzzle remains. Where does the light come from? There are no photons in the theory. Quantum mechanics offers a resolution. General principles of quantum mechanics require that any classical oscillator of frequency $\omega$ in its quantum version has quantum energy states of energy $E(n)=(n+\frac{1}{2})\hbar\omega,n=0,1,...$. The lowest energy state of the quantum system is no longer zero, as for a classical oscillator,  but is $\frac{1}{2}\hbar\omega$. This zero-point energy is a fundamental quantum mechanics concept. Now photons are described classically as fields which become operator fields in quantum theory. 
A classical field can be written as a Fourier transform of the space and time variables.
The time classical  Fourier transforms are integrals of the form $\int d\omega e^{i\omega t}f(\omega,\vec{x})$. They can thus we viewed as a collection of classical oscillators with all possible values for their frequencies $\omega$. In the quantum theory each oscillation frequencies will have  quantum energy $E(\omega)=(n+\frac{1}{2})\hbar\omega$. The lowest energy state of the system will then consist of of light quanta, photons, of all frequencies present, all as zero-point energies. In this picture photons are always virtually present. They are hidden in the zero-point energy of the photon field. Hence when energy is injected into the system photons can emerge. 

In quantum mechanics  self adjoint operators replace classical functions and the  real eigenvalues of these operators are identified as the possible allowed results of measurement. This immediately means that self adjoint operators that do not commute cannot be assigned exact numbers at the same time. There is thus a maximal set of commuting self adjoint operators  for a given system that can all be assigned well defined real numbers when measured that can be used to label the states of the system. Thus for the pair of operators, corresponding to the momentum and position vectors, a state can be described by either as an eigenstate labeled by the eigenvalue of position or of momentum since the two operators do not commute. We will discuss features of self adjoint operators more carefully later on. 

Soon after quantum mechanics was  proposed its methods were used to derive Balmer's formula. In quantum mechanics the concept of an orbit  no longer holds as the position and momenta labels, used to define an orbit, can no longer be assigned at the same time. The physical picture of hydrogen is  now very different.

We show how Balmer's expression can be obtained in two formulations of quantum mechanics: the operator approach of Heisenberg and the differential equation approach of Schroedinger. We also show how the Schroedinger differential equation can be obtained  from an operator equation. These historical results are necessary as they help us to unravel the subtle problem hidden in relating  the degeneracy of bound state energy levels to physically relevant variables. 

Pauli used the operator formulation of quantum mechanics to establish  properties of hydrogen atom\cite{pauli,bander}.  But we first  need to summarize some mathematical ideas  and facts about symmetries of circular orbits for planetary systems and establish properties of $\vec{A}$.
\subsubsection*{Symmetries of planetary orbits under $\frac{1}{r}$ potentials}  
It was discovered that a planet orbiting under an attractive $\frac{1}{r}$ potential 
has an extra conserved vector$\vec{A}$,  called the Laplace-Runge-Lenz (LRL) vector,  besides the angular momentum vector $\vec{L}$\cite{sub}. Since our aim is to show how this vector, in its quantum version, directly explains the degeneracy of energy eigenstates we need to establish properties of the LRL vector. We use a few ideas of differential geometry for curves in three dimensions to do this. After that we construct a quantum version of this vector and discuss Pauli's results.  
\subsubsection*{Basic Geometric Results: Frenet Frames}
Consider a  closed planetary orbit in three dimensional space. There is a standard
differential geometric way to describe such curves which we now  outline. Let us describe the curve as $\beta(s)$ in three dimensional space and use its arc length $s$
to parametrize it. Since the curve is in three dimensions three coordinates are needed to describe it. The idea of the Frenet  was to introduce three local coordinates axes fixed to a point on the curve and study the way they evolve when the parameter $s$ changes. The first coordinate axis he choose was  $T(s)$ the unit tangent vector to the curve defined by the equation,
\begin{eqnarray}
T(s)=\frac{d\beta(s)}{ds}
\end{eqnarray}
Where "s" is the small arc length of the curve. Since T is the unit tangent vector, we have,
\begin{eqnarray}
T.T=1
\end{eqnarray}
Taking derivative of this expression with respect to "s" we get the following:
 \begin{eqnarray}
T.T{'}=0\Rightarrow T \perp T^{'}.
\end{eqnarray}
we next choose as an axis  N  a unit vector orthogonal to T. This condition can be satisfied by choosing N to be parallel to $T^{`}$. We write,  
 \begin{eqnarray}
T^{'}(s)=kN(s)
\end{eqnarray}
where and k is a scalar which can be identified as the curvature of the curve and
N(s) is the unit normal. We have,
 \begin{eqnarray}
k=|| T^{'}(s)||
\end{eqnarray}
Finally we introduce our third vector that is orthogonal to T and N, given by,
 \begin{eqnarray}
\vec{B}=\vec{T}\times \vec{N}
\end{eqnarray}
which is called the binormal vector. We have the following results, 
\begin{eqnarray}
T.T=N.N=B.B=1;\\
T.N=T.B=N.B=0;
\end{eqnarray}
Thus $\vec T\perp\vec N\perp\vec B$\\
Now
\begin{eqnarray}
T.B=0\\
\end{eqnarray}
After taking a derivative wrt ''s"\\
\begin{eqnarray}
B^{'}.T=-T^{'}.B\\
\vec {B^{'}}.\vec {T}=-k\vec {N}.\vec {B}=0\\
\vec {B^{'}}.\vec {T}=0\\
\vec {B^{'}}\perp\vec {T}\\
\vec{B^{'}(s)}=-\tau \vec{N(s)}
\end{eqnarray}
Similarly we  get:\\
\begin{eqnarray}
\vec{N^{'}}=-k\vec {T}+\tau\vec{B}
\end{eqnarray}
We now have the differential geometry tools we need to study the LRL  vector for a planetary orbit. As planetary orbits are confined to a plane we only need to use the vectors $\vec{T},\vec{N}$. The binormal vector will not be required. 
\subsection*{LRL vector in Frenet frames}
For the planar gravitational Kepler problem we use polar coordinate system to describe the position of the particle in the plane. Then the velocity vector is given by,
 \begin{eqnarray}
\vec{v}=\dot{r}\hat{r}+r\dot{\theta}\hat{\theta}
\end{eqnarray}
This vector $\vec{v}$ always points in tangential direction to the orbit it traverses, it is thus the first of our Frenet vectors. We write it as, 
\begin{eqnarray}
\vec{T}=\frac{\vec{v}}{|\vec{v}|} 
\end{eqnarray}
We next find the Frenet vector $\vec{N}$, perpendicular to $\vec{T}$. This vector is   in the trajectory plane, and is given by, 
\begin{eqnarray}
\vec{N}=\frac{\dot{r}\hat{\theta}-r\dot{\theta}\hat{r}}{\sqrt{\dot{r}^2+r^2\dot{\theta}^2}}
\end{eqnarray}
Now first introduce the  LRL vector by the definition, 
\begin{eqnarray}
\vec{A}=\vec{p}\times\vec{L}-mk\hat{r}
\end{eqnarray}
We will shortly show that it is a conserved vector, that is, $\frac{d \vec{A}}{dt}=0$ follows from the equation of motion $\frac{d\vec{p}}{dt}=-km\frac{\vec{r}}{r^3}$. But our initial aim is to find the direction and magnitude of $\vec{A}$. We determine the direction by first expressing $\vec{A}$ in terms of the vectors $(\vec{T}, \vec{N})$. 
Using polar coordinates we get :
\begin{eqnarray}
\vec{A}=|\vec{L}||\vec{p}|\vec{N}-mk\hat{r}
\end{eqnarray}
Now using Eq.18,19 and 20.We can express $\hat{r}$ as a linear superposition of $\vec{T}$ and $\vec{N}$\\
\begin{eqnarray}
\hat{r}=\frac{\dot{r}}{|v|}\hat{T}-\frac{r\dot{\theta}}{|v|}\hat{N}
\end{eqnarray}
\begin{eqnarray}
\vec{A}=(|\vec{L}||\vec{p}|+\frac{mkr\dot{\theta}}{|v|})\vec{N}-mk\frac{\dot{r}}{|v|}\hat{T}
\end{eqnarray}
\subsection*{LRL conservation in -1/r potential}
We now show that $\vec{A}$ is conserved for a $\frac{1}{r}$ potential. Taking a time derivative  of $\vec{A}$ we get, 
\begin{eqnarray}
\frac{d\vec{A}}{dt}=\dot{\vec{p}}\times \vec{L}+ \vec{p}\times \dot{\vec{L}}-mk\frac{d\hat{r}}{dt}
\end{eqnarray}
Now in central force $\vec{F}=\dot{\vec{p}}=f(r)\hat{r}$ and $\dot{\vec{L}}=0$\\
Therefore the second term vanishes and first term simplifies:\\
\begin{eqnarray}
\dot{\vec{p}}\times \vec{L}=-mf(r)r^2\frac{d\hat{r}}{dt}
\end{eqnarray}
Now using Eq.32 and Eq.33 we note that only when $f(r)=\frac{-1}{r^2}$ ($\frac{-1}{r}$ potential)do  the first and last term cancels each other making $\vec{A}$ a conserved vector of the motion. 

For the planetary Kepler problem, with an inverse-square force law, we know that orbits are a conic section. They are ellipses when the eccentricity $e<1$. From Eq.23 we see that at the two turning points of the elliptic orbits, the perihelion and aphelion, the tangential component of velocity will be zero as $\dot{r}=0$ so that the LRL vector will always point in the line joining one focus to another in the direction of $\vec{N((\dot{r})}=0)=-\frac{\vec{ r}}{r}$. Thus the direction of $\vec{A}$ has been determined. We proceed to determine its magnitude.

First we set $m=k=1$. This means that we are choosing special units so that algebraic results obtained can be directly used to discuss planetary or electron motion under a $\frac{1}{r}$ potential. Thus we now write the classical LRL vector $\vec{A}$ as, 
\begin{displaymath}
 \vec{A}=(\vec{p}\times\vec{L})-\frac{\vec{r}}{r}
 \end{displaymath}
 where $r=|\vec{r}|=+\sqrt{\vec{r}.\vec{r}}$. It is then easy to show, using vector identities that,
 \begin{eqnarray}
 A^2=2EL^2+1
 \end{eqnarray}
 This fixes the magnitude of $\vec{A}$,  where $ A^2=\vec{A}.\vec{A},~ L^2=\vec{L}.\vec{L}$. A video of the way $\vec{A}$ remains unchanged while a
 planet goes round a circular orbit was constructed is attached and can be viewed in the following site.
\subsubsection*{The Quantum Hydrogen Atom} 
We now show that a quantum version of the LRL vector as a self adjoint operator exists.We construct this self adjoint operator by replacing the classical vector functions $\vec{p},\vec{L}, \vec{A}$ by appropriate self adjoint operators.  The self adjoint form for the operators  $(\vec{L}, \vec{r})$ are well known. Thus we only need to define a self adjoint version of $\vec{A}$.  
 
\subsubsection*{Quantum LRL Vector} 
We consider the self adjoint combination, chosen by Pauli,
 \begin{displaymath}
 \vec{A}=\frac{1}{2}[(\vec{p}\times\vec{L})-(\vec{L}\times\vec{p})]-\frac{\vec{r}}{r}
 \end{displaymath}
 and establish its  formal self adjoint property. Since the term $\frac{\vec{r}}{r}$ is self adjoint, we only need to check if the cross product term combination introduced is formally self adjoint.  We have 
 \begin{eqnarray*}
 (\vec{p}\times\vec{L})_i&=&\epsilon_{ijk}p_jL_k\\ 
(\vec{p}\times\vec{L})_i^{\dagger}&=&\epsilon_{ijk}L_kp_j\\
&=&-\epsilon{ijk}L_jp_k
\end{eqnarray*}
In addition we the following commutator relationship can be established,( units $\hbar=1$)
\begin{eqnarray}
 [L_i, L_j] &=& i\epsilon_{ijk}L_k\\
 {[L_i, V_j]}&=& i\epsilon_{ijk}V_k\\
 {[H, A_i]} &=& 0\\
 {[H, L_i]} &=& 0\\
 {[A_i, A_j]} &=& -2iH\epsilon_{ijk}L_k
 \end{eqnarray}
 where $H=\frac{\vec{p}^2}{2}-\frac{1}{r}$ is the Hamiltonian operator in our
 units. The vanishing of the commutator of $A_i$ and $L_i$ with the Hamiltonian $H$  imply that even in the quantum theory there is an additional symmetry, associated with $A_i$. Pauli used these commutation relations to find the eigenvalues
 of $H$ by first replacing the LRL vector operators $A_i$ by $D_i=\frac{A_i}{\sqrt{-2H}}$. Then the commutator $[A_i,A_j]=-2iH\epsilon_{ijk}L_k$ is replaced by  $[D_i,D_j]=i\epsilon_{ijk}L_k$ and the system of commutators has a Lie algebra structure. Next Pauli introduced two linear combinations of the operators $I_i=\frac{L_i+,D_i}{2}, K_i=\frac{L_i-D_i}{2}$ both of which commute with $H$ and satisfy the following commutation relations,
 \begin{eqnarray}
 [I_i,I_j]&=&i\epsilon_{ijk}I_k\\
 {[K_i,K_j]}&=&i\epsilon_{ijk}K_k\\
 {[I_i,K_j]}&=&0
 \end{eqnarray}
Thus the algebraic structure consists of two copies of angular momenta
operators that mutually commute. Hidden in these operators is the formal operator 
$\frac{1}{\sqrt{-2H}}$ that contains the Hamiltonian operator $H$. The eigenstates
of the formal objects $I_i,K_i$ are known, since $H$ commutes with both $I_i,K_i$
these eigenstates can also be eigenstates of $H$. By switching to
eigenstates of $I_i,K_i$ operators we show how Pauli was able to determine the eigenvalues of $H$ but there are surprises. 

We observe that $I^2=K^2$. The eigenstates of the system with eigenvalues are,
\begin{eqnarray}
I^2|i,i_z:k,k_z>&=&i(i+1)|i,i_z:k,k_z>\\
I_z|i,i_z:k,k_z>&=&i_z|i,i_z,k,k_z>\\
K^2|i,i_z;k,k_z>&=&k(k+1)|i,i_z;k,k_z>\\
K_z|i,i_z;k,k_z>&=&k_z|i,i_z;k,k_z>
\end{eqnarray}
These states are also energy eigenstates since $H$ commutes with $I^2,K^2,I_z,K_z$ . In general $i,k$ can range over either integer values $0,1,2,...$ or half integer values $\frac{1}{2},\frac{3}{2},...$. But as $I^2=K^2$ it follows that we must set $i=k$. We note that there are two disjoint ranges of values allowed for $(i,k)$, either both take equal integer values or both take half integer values. Finally for the quantum case
the expression for $A^2$ is,
\begin{eqnarray}
\frac{1}{-2H}&=&\frac{A^2}{2|H|}+L^2+1\\
&=&D^2+L^2+1\\
D^2&=&I^2+K^2-2\vec{I}.\vec{K}\\
L^2&=&I^2+K^2+2\vec{I}.\vec{K}\\
\frac{1}{-2H}&=&4I^2+1
\end{eqnarray}
 Thus we have the energy eigenvalue, 
\begin{displaymath}
\frac{1}{-2E_n}=4i(i+1)+1=(2i+1)^2= (m+1)^2=n^2
\end{displaymath}
where we have used the fact that $I^2=K^2$. 

We now have our first surprise.  To reproduce the Bohr result, we must have $n=1,2,...$ This requires that $2i=0,1,2,...(n-1)$ but for this $ i$ must range over both integer as well as half integer values. This is a surprise.  We next note that the expression for $E_n$ is independent of both $i_z$ and $k_z$. Thus the energy eigenstates  have degeneracy that reflects this property.
This degeneracy $D(E_n)$  can be calculated. We have $d(E_n)=\sum_{m}(2m+1), m=0,1,..(n-1)$ .This gives $d(E_n)=n^2$. In the calculation  we have used the fact that the number of states $i_z$ for a given integer value $i$ is $d_0=(2i+1)$ but when $i$ is a half integer the number of states is $d_1=2i$. Since these appear disjointly we have $d=d_0+d_1=4i+1$. The degeneracy due to for $k_z$  has also been taken into account as we have used the result $I^2=K^2$ as well. Thus $d(E_n)=(4i+1)=(2m+1)$. This is the correct degeneracy for hydrogen levels. 

Thus our analysis has uncovered a surprising feature of Pauli's approach. By
including the LRL vector and introducing the square root of Hamiltonian operator to define $D_i$, an algebraic system that reflects the symmetry of the hydrogen Hamiltonian was generated. In this abstract setting the underlying symmetry was captured by two sets of angular momentum operators which had both integer as well as half integer angular momentum values.  Once this step was taken the abstract system showed that the spectrum of the hydrogen atom was precisely that obtained by Bohr and that this spectrum was degenerate with the degree of degeneracy fixed by the symmetry. However the a physical interpretation of $n$ as well as the nature of the degenerate states was lacking.  The next natural question is: can Pauli's approach directly give the hydrogen wave functions? The answer is no\footnote{An indirect method of finding the wave functions from the Pauli eigenstates is possible\cite{chua}}. The physical hydrogen spectrum is determined separately by solving the eigenvalue equation $H|n,l,l_z>=-\frac{1}{2n^2}|n,l,l_z>, n=1,2,..,l=0,1,...(n-1), -l\leq l_z \leq+l$. The solution shows that the principal quantum number,$n$, determines the spatial extent of the wave function, while $l,l_z$ represent the orbital angular momentum that can only have integer values. In the approach of Pauli, on the other hand, the spectrum is determined ,not from the space-time physics but from an abstract mathematical symmetry of the hydrogen Hamiltonian. Consequently the physical nature of the degeneracy is not revealed. It is simply a consequence of an abstract symmetry, described in terms of two angular momentum operators, both of which must be allowed to have integer as well as half integer values. These mathematical angular momenta do not reflect the properties of the physical hydrogen atom in three dimensional space where only integer orbital angular momentum values are allowed.  

One further technical remark can be made. We note that the square of  LRL vector, $D^2=D^{\dagger}D$, where $D^{\dagger}$ is the adjoint of $D$,has eigenvalues
has an expectation value $<D^2>= i(i+1)+k(k+1)-2i_zk_z>0$ and thus $D^2\geq0$. It is thus  consistent with the fact that $D_i$ is a self adjoint operator. We will see that this is not the case when $D^2$ is evaluated on the space of the eigenstates of $H$ in the Schroedinger wave mechanics approach. This result reveals subtle features present in the concept of self adjointness that we will comment on later.
      
We would like to identify $A^2$ as an extra quantum number label for the eigenstates of $H$ and show that it explicitly helps us understand the degeneracy of the eigenstates found by calculations. As we pointed out that this step cannot be taken in the framework of Pauli, where the eigenstates of $H$ used do not have physically relevant eigenvalue labels. We need to work out $A^2$ for the physical eigenstates of $H$. Thus our first step is to find an expression for $A^2$ in terms of operators that have well defined eigenvalues for such eigenstates of $H$.  Let us give some of the algebraic details required to do this. We have,
 \begin{eqnarray}
 (\vec{p}\times\vec{L})^2&=&p^2L^2\\
 (\vec{L}\times\vec{p})^2&=&L^2p^2\\
 (\vec{L}\times\vec{p}).(\vec{p}\times\vec{L})&=&-p^2L^2\\
 (\vec{p}\times\vec{L}).(\vec{L}\times\vec{p})&=&-4p^2-p^2L^2
 \end{eqnarray}
 The analogous results for the four terms of the form $\frac{1}{r}(\vec{r}.(\vec{p}\times\vec{L})$, with different factor orderings, are,
 \begin{eqnarray}
 \frac{\vec{r}}{r}.(\vec{p}\times\vec{L})&=&\frac{L^2}{r}\\
 \frac{\vec{r}}{r}.(\vec{L}\times\vec{p})&=&-\frac{L^2}{r} +2i\frac{\vec{r}}{r}.\vec{p}\\
 (\vec{p}\times\vec{L}).\frac{\vec{r}}{r}&=&\frac{L^2}{r}+2i\vec{p}.\frac{\vec{r}}{r}\\
 (\vec{L}\times\vec{p}).\frac{\vec{r}}{r}&=&-\frac{L^2}{r}
 \end{eqnarray} 
 The Hamiltonian of our three dimensional system is  defined $H=\frac{\vec{p}^2}{2}-\frac{1}{r}$, where $r=|\vec{r}|$. In classical mechanics case $p^2$ is the square of the classical momentum vector while for the quantum case it is the the square of the corresponding vector momentum operator. With the help of these results we get,
 \begin{eqnarray}
 A^2=2H[L^2+1]+1
 \end{eqnarray}
 Notice the presence of an extra term $H$ in the quantum expression for $A^2$
 compared to the classical expression obtained earlier. Our next step is to find physical eigenstates of $H$.
 \subsubsection*{The Factorization Method}
 We use a well known algebraic method for determining the bound state spectrum for hydrogen and its eigenstates that have physically relevant eigenvalue labels. The special features of this algebraic approach are that it throws light on the physical meaning of the principal quantum number and it explicitly shows that the bound state energy levels of hydrogen do not depend on the eigenvalues of the $L^2=\vec{L}.\vec{L}$  angular momentum operator $\vec{L}$. 
 
The factorization method  starts with the with radial eigenvalue differential Schroedinger equation for a hydrogen atom bound eigenstate and solves it using an algebraic method in a manner similar to the raising and lowering operator method used by Dirac\cite{dirac} to solve the harmonic oscillator problem. Thus once  the physical eigenvalues are found we can express the LRL vector in terms of the quantum numbers that correspond to a given energy eigenvalue state and show how it provides a clear physical picture of the degeneracy of eigenstates. But as we will see there is  mathematical problems lurking in the background. 
 
 The  radial eigenvalue equation for hydrogen atom is obtained from the Schroedinger  equation in a few steps. We start with the operator equation 
 \begin{displaymath}
  H|\psi>=[\frac{\vec{p_{op}}^2}{2m}-\frac{e^2}{\sqrt{\vec{r_{op}}.\vec{r_{op}}}}]|\psi>=-E|\psi>
 \end{displaymath}
 where the operators satisfy the commutation rules $[p^{op}_i, r^{op}_j]=-i\delta_{ij}\hbar$, where $\delta_{ij}=1,i=j$, is zero otherwise and $[p^{op}_i,p^{op}_j]=[x^{op}_i,x^{op}_j]=0$ and can be represented as follows, 
\begin{eqnarray*}  
  \vec{p_{op}}&\rightarrow& -i\hbar\nabla\\
\vec{r_{op}}&\rightarrow &\vec{r}\\
|\psi>&=&\psi(\vec{r})
\end{eqnarray*}
Using this representation the operator equation is converted  to the Schroedinger differential equation,
 \begin{displaymath}
 [-\frac{\hbar^2}{2m}\nabla^2-\frac{e^2}{r}]\psi(\vec{r})=-E\psi(\vec{r})
 \end{displaymath}
 where $e$ is the electron charge, $m$ the electron mass and $\vec{r}$ is the distance of the electron from the central proton of charge $+e$.We then rewriting this equation using spherical polar coordinates $(r,\theta,\phi)$ with ranges $( 0\leq r\leq \infty, 0\leq \theta\leq \pi,0\leq \phi \leq 2\pi )$ and then set $\psi(\vec{r})=R_l(r)Y_{lm}(\theta,\phi)$, where $Y_{lm}(\theta,\phi)$ are spherical harmonics, labeled by integers $(l,m)$ that have ranges: $(l=0,1,... ),(-l \leq m\leq +l)$. These harmonics
 can be used to represent any arbitrary square integrable function  $f(\theta,\phi)$. We  will replace  the label $m$ by the label $l_z$ later on. Next we use the fact that the operator $\nabla^2$ in spherical polar coordinates can be written as,
 \begin{eqnarray*}
 \nabla^2&=&\frac{1}{r^2}\frac{\partial}{\partial r}(r^2\frac{\partial}{\partial r})+\frac{L^2}{r^2}\\
 L^2&=&\frac{1}{\sin\theta}\frac{\partial}{\partial \theta}(\sin\theta\frac{\partial}{\partial \theta})+\frac{1}{\sin\theta^2}\frac{\partial^2}{\partial^2 \phi}
 \end{eqnarray*}
and  $ L^2Y_{lm}(\theta,\phi)=-l(l+1)Y_{lm}(\theta,\phi)$ to get the radial equation,
  \begin{displaymath}
- \frac{1}{r^2}\frac{d}{dr}(r^2\frac{dR_l}{dr})+\frac{l(l+1)}{r^2}R_l=-(\frac{2 m}{\hbar^2})(E_l+\frac{e^2}{r})R_l(r)=0
 \end{displaymath}
Finally we write $\Phi_l(r)=\frac{R_l(r)}{r}$, we get the radial Schroedinger equation in the form that we need, namely,
\begin{displaymath}
H_l\Phi_l=[-\frac{d^2}{d\rho^2}-\frac{2Z}{\rho}+\frac{l(l+1)}{\rho^2}]\Phi_l=\epsilon_l\Phi
\end{displaymath}
where we have introduced dimensionless variables,  $\rho=\lambda r, \lambda =\frac{me^2}{\hbar^2}$ and $\epsilon=\frac{2\hbar^2}{me^4}E<0 $. The equation is now  dimensionless. The wave function $\Phi$ is normalized using the one dimensional volume, that is $\int_{0}^{\infty}d\rho |\Phi|^2 =1$. The parameter $l$ can be zero or any positive integer.

 We now factorize the second order radial differential operators for $H_l$ \cite{infeld, schroedinger}as a product of two linear first order differential operators, $h_l,h^{\dagger}_l$. The idea is to use the algebraic commutation properties of these operators to determine the spectrum of the quantum hydrogen atom  following steps similar to those used in the ladder operator approach to find the energy levels of the harmonic oscillator problem. We introduce the following operators:
\begin{eqnarray}
h_l&=&i(\frac{d}{d\rho}+V_l(\rho))\\
h_l^{\dagger}&=&i(\frac{d}{d\rho}-V_l(\rho))
\end{eqnarray}
where $V_l(\rho)=(\frac{Z}{l+1}-\frac{l+1}{\rho})$. Then
\begin{eqnarray}
h_l^{\dagger}h_l&=&H_l+\frac{Z^2}{(l+1)^2}\\
h_l h_l^{\dagger}&=&H_{l+1}+\frac{Z^2}{(l+1)^2}
\end{eqnarray}
and, 
\begin{eqnarray}
[h_l,h^{\dagger}_l]&=&+\frac{2(l+1)}{\rho^2}\\
h_l H_l&=&H_{(l+1)}h_l
\end{eqnarray}
where $[a,b]=ab-ba$, the commutation relation. These relations imply that if $E_l$ is an eigenvalue of $H_l$ then $E_{(l+1)}$ is an eigenvalue of $H_{(l+1)}$ where $|E_{(l+1)}>=h^{\dagger}_l|E_l>$. This means acting on an eigenstate with the operator $h^{\dagger}_l$ increases the value of $l\rightarrow (l+1)$. We require $E_l<0$ in order to have bound states but increasing $l$ increases the energy.  Thus for a given energy eigenvalue $E_L$, there must be a state $|E_L>$ such that $h_l|E_L>=0$. This is the essential step.

  An intuitive way to understand this requirement  is to plot the effective potential  $U(\rho)=-\frac{Z}{\rho}+\frac{l(l+1)}{\rho^2}$ as a function of $\rho$ and drawing a line parallel to the $\rho$ axis to represent an energy eigenvalue $E$. For $E<0$, the line intersects the $U(\rho)$ at two points. These points represent the turning points of motion in a classical sense. The distance between them defines the classical orbit size. Here our condition $E<0$ implies a maximum allowed angular momentum value $L$, with the property that if  $L\rightarrow(L+1)$ the $E_L$ becomes positive.  This intuitive idea is implemented by the constraint $h_L|E_L>=0$ for all integer including the value zero, for $l=L$. In terms of the effective potential, the turning  points are now separated by the largest distance possible for a given integer $L$ and $E_L<0$. The corresponding size of the quantum state can be taken to be the expectation value $<E_L|\frac{1}{r}|E_L>$ in that eigenstate.  

\begin{center}
		\includegraphics[scale=0.8]{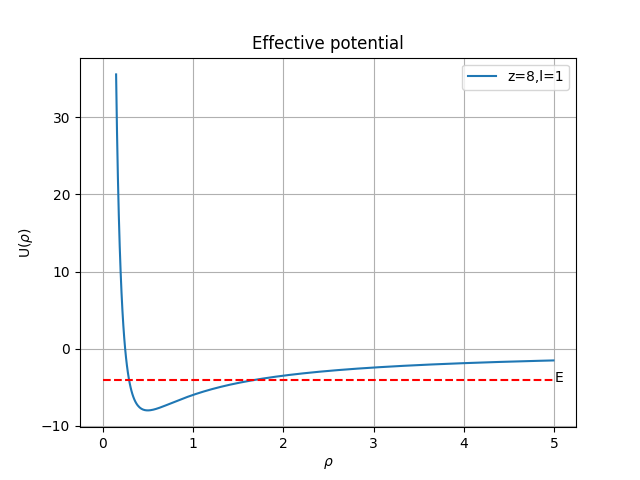}
		Fig 1:Effective potential 
	\end{center}
The classical effective potential picture identifies a conceptual problem. If we set $l=0$ the potential is always negative and has no lower bound. What is the reason 
for the lowest energy eigenvalue to be bounded?  Quantum mechanics provides a clear reason. It is because $\vec{p},\vec{x}$ are operators in quantum mechanics that can fluctuate. By represented the operators $\vec{p_{op}}\rightarrow -\hbar\nabla, \vec{x_{op}}\rightarrow \vec{x}$  where we represent eigenstates as wave functions $\psi(\vec{x},t)$ we will later on give a general quantum argument, which shows that  $-E=<H>$ is bounded. It must be greater than $-E_0$ where $E_0$ is finite. This means the electron cannot collapse to the center. An essential step in the argument is requiring the eigenstates to be normalizable, that is to have a scalar product $<E_L|E_L>=1$.

 Let us now complete our calculation of the eigenvalue. We set $h_L|E_L.=0$, act on this state with $h^{\dagger}_L$ and use the commutation relations, to get,
\begin{eqnarray}
h^{\dagger}_Lh_L|E_L>&=&[H_L+\frac{Z^2}{(L+1)^2}]|E_L>\\
&=&0
\end{eqnarray}
Hence we have found  the eigenvalue of $H_L$. It is $E_L=-\frac{Z^2}{(L+1)^2}$. The value of $L$ is a positive integer. This result shows that the principal quantum number is related to the expectation value of $\frac{1}{r}$. It is a length scale associated with an energy eigenstate that indicates the degree of localization of a given bound electron. The corresponding eigenfunction $\psi_{L}(\rho) $ can also be easily determined by solving the equation,
\begin{displaymath}
h_L\psi_{L}(\rho)=i(\frac{d}{d\rho}+V(\rho))\psi_{L}(\rho)=0
\end{displaymath}
which gives, $\psi_{l}(\rho)=K\rho^{L+1}e^{-\frac{Z\rho}{L+1}}$ where
$K$ is a normalization constant.  This eigenstate is known to be degenerate, that is all lower values $L=l$ have the same energy. This follows from  Eq(6) since it  implies that,
\begin{displaymath}
H_{L}=h_{L-1} H_{L-1}(h_{L-1})^{\dagger}
\end{displaymath}
so that $H_L|E_L>=E_L|E_L>=h_{L-1}H_{L-1}(h_{L-1})^{-1}|E_L>$ which implies $H_{L-1}(h_{L-1}^\dagger|E_L>)=E_L(h_{L-1}^\dagger|E_L>)$. Thus a state with eigenvalue $E_L$ is degenerate. By iterating this procedure it follows that all lower values of $L$ do not change the eigenvalue $E_L$. We can thus write the eigenvalue as $E(n,l)=-\frac{Z^2}{n^2}, n=(L+1), l=0,1,...,L$, $n$ is the principal quantum number and also rotational invariance of $H$ results in $E(n,l)$ being independent of $l_z$ values that have the range $(-l \leq l_z \leq +l)$. Thus the energy eigenstates carry three labels, namely, $(n,l,l_z)$ but the energy eigenvalues do not depend on $(l,l_z)$. The labels $(l,l_z)$ are physical labels that describe the degenerate states
and we have identified the principal quantum number $n=1,2,..$ as a measure of the size of an eigenstate, since $<\frac{1}{r}>=\frac{1}{n^2a_0}$ where  $a_0=\frac{me^2}{\hbar^2}$

Our next step is to write $A^2$ in terms of these physical eigenvalue labels of the hydrogen atom Hamiltonian. The idea is to then use $A^2$ as an additional label for an energy eigenstate. This is possible as $A^2$ commutes with $H,L^2,l_z$. Once this is done we can identify $A^2$ as a degeneracy label since its value can change without changing the energy eigenvalue.  Our aim to relate the LRL vector explicitly to
to the degeneracy of eigenstates would be achieved. But when we do this we run into a problem which we now  explain.

  Let us first write down the expression for the expectation value of  $A^2$ in terms of the eigenvalues of $L^2, L_z$ and $H_l$, using the expression for $A^2$ obtained earlier. We have, 
 \begin{eqnarray}
 A^2+1&=&-\frac{2[l(l+1)+1]}{(L+1)^2}+2\\
 &=&\frac{-2[l(l+1)+1]+2(L+1)^2}{(L+1)^2}
  \end{eqnarray}
The expression shows that the expectation value of operator $\vec{A}^2+1\geq0$. Indeed for $L=l=0$ we have $A^2+1=0$, thus $A^2$ can be negative. This is the problem. Since $A^2$ is the square of a  formally self adjoint operator all of its eigenvalues should be positive. This problem needs to be resolved. A simple way of doing this is to replace  $\vec{A}$ t by $\vec{B}_{L,l}$  defined to be: 
 \begin{eqnarray}
 \vec{B_{L,l}}&=&\vec{A}+\frac{l_z}{|l_z|} ~~\mbox{for all}~~ l\\
\frac{l_z}{|l_z|}&=&1 ~~\mbox{for}~~l_z=0
\end{eqnarray}   
 This vector also commutes with the Hamiltonian $H$ and $L^2$  is thus an independent conserved vector of motion. It is a modified LRL vector, $B^2=A^2+1$,  and now $B^2\geq 0$ its eigenvalues are never negative. Thus the problem of negative eigenvalues has been resolved by introducing  the vector $\vec{B}$. We now have, 
 \begin{displaymath}
 \frac{B^2}{2E(n)}+(l(l+1)+1)=\frac{1}{E(n)}
 \end{displaymath}
 with $n=(L+1), l=0,1,..L$. As $l,l_z$ vary $E(n)$ remains unchanged. Thus the 
 combination $\frac{B^2}{2E(n)}+(l(l+1)+1)$ remains the same, where the additional factor of one comes from the unit vectors $\hat{l_z}^2=1$ for each $\hat{l_z}$ value
 including $l_z=0$. 
 
The number of degenerate  states  corresponding to a given energy eigenvalue
can now be explicitly related to the eigenvalues of  $B^2,L^2,L_z$. The key observation is that our expression for $B^2$ shows that these eigenvalues can all vary keeping $E(n)$ the same.  Thus the number of degenerate of eigenstates  can be determined by simply counting range of  values of  that $(l,l_z)$ can take, for a fixed energy state $E(n), n=(L+1)$ since for any such state the $B^2$ eigenvalue can be adjusted to keep the value of $E(n)$ the same. Thus since  $l$ ranges over the values $l=0,1,...L$ and for each $l$ value $l_z$ range from  $-l\leq l_z \leq +l$ the degeneracy of and energy $E(n)$ eigenstates states, $d(E(n))$ is given by
\begin{displaymath}
d(E(n))=\sum_{l=0}^{(n-1)}(2l+1)=n^2
\end{displaymath}
 We have thus been able to directly relate degeneracy of energy eigenstates to the presence of the conserved vectors of $H$. However to do this we had to modify the LRL vector required to make it self-adjoint  in the space of the physical eigenstates of the hydrogen atom Hamiltonian $H$.
 
 \subsubsection*{Discussion}
 Our focus was to understand how the degeneracy of the bound state energy eigenvalues  of hydrogen atom Hamiltonian $H$ is related to its symmetries in an explicit way. We found that this could be done but it required resolving a problem regarding the self adjoint property of the LRL vector on the eigenstates of $H$.
 
 We  now point out some conceptual aspects of the quantum mechanical model. There is no electromagnetic radiation for electrons in the eigenstates of $H$, since the expectation value of the acceleration operator in an eigenstate is zero. This is easily 
 established by a  calculation. The acceleration operator, $\vec{a_{op}}=\frac{d \vec{p_{op}}}{dt}$, where $\vec{p_{op}}$ is the momentum operator. The expectation value of this operator for an eigenfunction, $\phi(\vec{x})$ of hydrogen atom is,
 \begin{displaymath}
 (\phi,\vec{a}\phi)=(\phi,\frac{i}{\hbar}[H,\vec{p}] \phi)=0
 \end{displaymath}
Thus the collapse of the electron to the center due to radiation of light is not an issue.

Quantum theory also does more, it provides a general argument for the stability of the lowest energy eigenstate $-E_0$ against a collapse to the center, by establishing a bound  for it $-E_0>-\epsilon_0$ which we now outline. This requires using a refined version of the uncertainty principle\cite{faris} and shows that the electron does not collapse to the center. It has stability. The most general argument does not require determining the eigenvalue and the result is space dimension dependent. The reason for such a dependence is that the Hamiltonian $H$ written down is valid for all dimensions. However, we use a less general approach and  restrict ourselves to $d=3$. In this dimension we use the following result of Hardy\cite{Hardy}, to find our bound for the energy eigenvalue,\\
\small{\textbf{Hardy Inequality}}
\begin{displaymath}
<p^2>\geq \frac{1}{4}<\frac{1}{|x|^2}>
\end{displaymath}
where we set $\hbar=1$ and$<A_{op}>=\int d^3x\psi^{*}(\vec{x})A_{op}\psi(\vec{x})$ for an operator $A$ acting on square integrable functions $\psi(\vec{x})$ with $|x|=\sqrt{\vec{x}.\vec{x}}$ and $p^2=\vec{p}.\vec{p}, \vec{p_{op}}\rightarrow-i\nabla, \vec{x_{op}}\rightarrow \vec{x}$, where $\vec{x_{op}}, \vec{p_{op}}$ are position and momentum operators, represented as described.The inequality is strict and the factor $\frac{1}{4}$ cannot be replaced by a smaller number. We next note that the hydrogen energy eigenvalue  can be written as,
\begin{displaymath}
E=<H>=<p^2>-\kappa<\frac{1}{|x|}>
\end{displaymath}
in suitable units. Using the Hardy inequality we thus get,
\begin{displaymath}
E>\frac{1}{4}<\frac{1}{|x|^2}>-\kappa<\frac{1}{|x|}>
\end{displaymath}
minimizing gives $E>-\kappa^2$. Thus the energy is bounded and electron wave function can be assigned a non-zero size. Putting back units the bound obtained can be written as $ E>-\frac{me^4}{\hbar^2}$, where $m, e$ are the electron mass and charge.
 
Since our calculations revealed subtle aspects of what it means for an operator to be self adjoint we add a few remarks clarifying the definition of  a self adjoint operator \cite{kumar}. If $(\phi,\psi)$ are elements 
 of a Hilbert space then an operator, $T$, is called symmetric if 
 \begin{displaymath}
 (\phi,T\psi)=(T\phi,\psi)
 \end{displaymath}
 where $(\phi,\psi)$ is the scalar product. Usually this relation is used to define a self adjoint operator. But  strictly this does not define a self adjoint operator. To properly define a self adjoint operator we first consider an operator $T$ and its adjoint $T^{*}$. These operators act on a subset of the functions present in Hilbert space selected by boundary conditions. This subset of functions is the domain of the operator.  Thus we write $D(T)$ as the domain of $T$ and $D(T^{*})$ as the domain of $T^{*}$. Given the domain $D(T)$ the domain $D(T^{*})$ is defined by requiring to
 find all elements $\phi$ of the Hilbert space so that
 \begin{displaymath}
 (\phi,T\psi)=(T^{*}\psi,\phi)
 \end{displaymath}
 The set of all such $\phi$ defines the domain of the adjoint $D(T^{*})$. This domain  need not be equal to $D(T)$. If $D(T)=D(T^{*})$ and $T=T^{*}$ then the operator is  self adjoint. The key point is the domains of an operator and its adjoint need not be the same. von Neumann proved a simple way to check if an operator is self adjoint. For a given symmetric operator $T$ check if there are solutions to the equations,
 \begin{eqnarray*}
 T\phi_{+}&=& +i\phi_{+}\\
 T\phi_{-}&=&-i\phi_{-}
 \end{eqnarray*}
 and let $(n_{\pm})$ be the number of square integrable linearly independent solutions with eigenvalues $i\pm$ respectively. Then $T$ is self adjoint or can be made self adjoint  if $n_{\pm}=0$ or if $ n_{+}=n_{-}$ if these conditions do not hold then $T$ is not self adjoint. It follows from this result that if $T^2\geq0$ and $T$ is a symmetric operator it is self adjoint. We used this result.  We also pointed out that $A^2\geq0$ for the eigenvectors used by Pauli. But although we have $(A\phi,A\psi)=(\phi,A\psi), A^2$ could be negative when it acted on the eigenstates of hydrogen atom.  Thus revealing a subtle feature of a self adjoint operator , namely that the same operator can be self adjoint for a certain set  of eigenstates but not for a different set. 
 
 A simple example illustrating von Neumann's theorem  might be helpful. Consider the one dimensional momentum operator  $P=-i\frac{d}{dx}$ on the half line
  $(0\leq  x\leq \infty)$. It is not a self adjoint operator. We check that on a square integrable function $\psi(x)$ the operator $P$ can have an imaginary eigenvalue.
  \begin{eqnarray*}
  -i\frac{d\psi}{dx}&=&i\psi(x)~~\mbox{has a solution}\\
  \psi(x)&=&\psi(0)e^{-x}~~(0\leq x\leq\infty)
  \end{eqnarray*}
 But there is no square integrable function on the half line with $-i$ eigenvalue.
 Thus $n_{+}=1,n_{-}=0$ and since they are not equal von Neuman's theorem
 tells us that the operator in this space is not self adjoint. 
    
Finally we pointed out conceptual features in Pauli's calculation. The main idea was to embed $H$, with an energy eigenvalue $E<0$, in a larger system that included the conserved vectors, $(\vec{A},\vec{L})$. The allowed eigenvalues and their degeneracy of $H$ then followed by requiring consistency of the  hydrogen bound state spectrum with this larger structure. The calculations made essential use of half integer values for angular  momentum states to label the relevant eigenstates. Thus the eigenvectors, with spin-like labels were not related to the eigenvectors of hydrogen atom and hence no direct understanding of the physical nature of the spectral degeneracy was possible.

The self adjointness of the  LRL operator introduced has also been studied by Thirring\cite{thirring} who showed that it was essentially self adjoint in the domain that contains the span of Hermite functions. We have showed that LRL in quantum theory is also self adjoint in the space of Pauli eigenstates of Hydrogen but that a modified LRL is self adjoint for the hydrogen wave functions. This is done by noting that in both cases a complete set of eigenfunctions exist and the square of the relevant operator $A^2\geq 0$. Further insights in this direction have emerged. For example a recent approach for solving the hydrogen atom problem,using ideas  reduction and unfolding, that use the quantum LRL vector and finds a link between a family of harmonic oscillators and the spectrum of hydrogen\cite{marmo}. The LRL vector approach has also been used to study the hydrogen atom problem for non-commuting space\cite{pres}.
\subsubsection*{Acknowledgement}
	Akshay Pal would like to thank Prof.Siddhartha Sen for providing him the opportunity to work on the problem and Prof. Jayanta K.Bhattacharjee for comments. He is very grateful to my friend Rahul Devarakonda for providing the visual simulation of the LRL vector and Shafaq Elahi for correcting his calculation mistakes in the factorization section.

\end{document}